\documentclass[aps,prc,preprint,groupedaddress]{revtex4}

\usepackage{graphicx}

\begin{document}


\title{Asymmetry of recoil protons in neutron beta-decay}


\author{V. Gudkov }
\email[gudkov@sc.edu]{}
\affiliation{Department of Physics and Astronomy, University of South Carolina,
Columbia, SC 29208 }


\date{\today}

\begin{abstract}
A complete analysis of proton recoil asymmetry in neutron decay in
the first order of radiative and recoil corrections is presented.
The possible contributions  from new physics are calculated in terms
of low energy coupling constants, and the sensitivity of the
measured asymmetry to models beyond the Standard model are
discussed.
\end{abstract}


\maketitle

\section{Introduction}

The  free neutron decay, being one of the simplest semi-leptotic
hadron decay processes,  is very important  in the search for
possible manifestations of new physics. The main advantage of
neutron decay is the possibility to describe the process with
minimal theoretical uncertainties and, as a consequence,  the
possibility to interpret unambiguously  experimental results. The
set of
 experiments for measurements of the neutron lifetime and
neutron decay correlations can be used to determine the weak vector
coupling constant, to test  the universality of the weak
interaction, and to search for  nonstandard couplings (see, for
example, \cite{gtw2,holsttr,deutsch,abele,yeroz,sg,herc,marc02} and
references therein). The detailed analysis of required experimental
accuracy and sensitivity to new physics of different observables for
standard setups in neutron decay experiments have been done in the
paper \cite{ggc}. However, the recent measurement \cite{CILL}  and
the new proposal to measure \cite{panda}  the integrated asymmetry
of recoiled protons in relation to the direction of  neutron spin
(which is known as a $C$-angular correlation coefficient
\cite{tr58,gl96}) raise the question about the sensitivity of this
asymmetry to new physics.  In order to be able to estimate the
potential sensitivity of the $C$ asymmetry to new physics and the
best accuracy of the measurement of Standard Model parameters (e.g.,
the ratio of axial-vector and vector coupling constants of weak
interaction),  one needs to calculate recoil and radiative
corrections for the $C$-asymmetry, as well as all possible
contributions from the model beyond the standard one. Moreover, all
these calculations must be done in the same framework to keep all
possible uncertainties under control.

In this paper, we use results of the effective field theory
description of neutron beta-decay \cite{eftcor} as a framework for
the calculation of the $C$ correlation coefficient in the Standard
Model (with recoil and radiative corrections). Then we calculate
possible corrections from new physics using the most general
non-standard beta-decay interactions. This  provides a consistent
description of the proton recoil asymmetry in terms of low energy
coupling constants related to models beyond the Standard one at a
level well below that anticipated in the next generation of neutron
decay experiments.

\section{Proton asymmetry in the Standard model }

 We have chosen results, based on the effective field theory (EFT)
 approach,
   of the  description of the polarized neutron decay
 since this approach provides a general expression for neutron decay distribution
function with the accuracy of $10^{-5}$ in terms of one free
parameter - low energy constant (LEC) (for more details, see paper
\cite{eftcor}). To calculate  the angular correlation coefficient
$C$ with the complete set of recoil and radiative corrections, we
use a general expression for the differential neutron decay rate
given by Eq.(8) in \cite{eftcor}. It should be mentioned, that in
the tree approximation (neglecting recoil corrections and radiative
corrections), the EFT results reproduce exactly well known formula
for neutron decay rate \cite{gtw1}  in terms of the angular
correlations coefficients $a$, $A$, and $B$:
 \begin{eqnarray}
\frac{d\Gamma ^3}{dE_ed\Omega_ed\Omega_{\nu}}= \Phi (E_e)G_F^2
|V_{ud}|^2 (1+3\lambda^2)
\hskip 2cm \nonumber \\
\times (1+b\frac{m_e}{E_e}+a\frac{\vec{p}_e\cdot \vec{p}_{\nu}}{E_e
E_{\nu}}+A\frac{\vec{\sigma} \cdot \vec{p}_e}{E_e}
+B\frac{\vec{\sigma} \cdot \vec{p}_{\nu}}{E_{\nu}}), \label{cor}
\end{eqnarray}
Here,  $\vec{\sigma}$ is the neutron spin; $m_e$ is the electron
mass, $E_e$, $E_{\nu}$, $\vec{p}_e$, and $\vec{p}_{\nu}$ are the
energies and momenta of the electron and antineutrino, respectively;
and $G_F$ is the Fermi constant of the weak interaction (obtained
from the $\mu$-decay rate). The function $\Phi (E_e)$ includes
normalization constants, phase-space factors, and standard Coulomb
corrections. For the Standard model the angular coefficients depend
only on one parameter $\lambda = -C_A/C_V >0$,  the ratio of
axial-vector to vector nucleon coupling constant (in general,
$C_V=C^\prime_V$ and $C_A=C^\prime_A$ are low energy coupling
constants for the low energy effective Hamiltonian given by
Eq.(\ref{ham})):
\begin{equation}
a=\frac{1-\lambda ^2}{1+3\lambda ^2},  \hskip 1cm A= -2\frac{\lambda
^2-{\lambda}}{1+3\lambda ^2}, \hskip 1cm B= 2\frac{\lambda
^2+{\lambda}}{1+3\lambda ^2}. \label{coef}
\end{equation}
 (The parameter $b$ is equal to zero for  vector - axial-vector  weak interactions.)

 The $C$ angular coefficient (do not mix with $C_V$ and $C_A$) has been defined \cite{tr58} as
the angular distribution of the recoil protons in the relation to
the direction of the neutron spin, provided all other variables,
including proton recoil momentum, are averaged out. In the tree
approximation, it has been calculated  in papers \cite{tr58,gl96},
and numerical corrections to this approximation have been calculated
in the paper \cite{gl96}. Using this definition, one can calculate
$C$-coefficient from a general expression for the differential
neutron decay rate (Eqs. (8)-(19) in  paper \cite{eftcor}) with all
(in the first order) recoil and radiative corrections. To do this,
we use the momentum conservation condition
$\vec{p}_{\nu}+\vec{p}_{e}+\vec{p}_{p}=0$ which is multiplied  by
the neutron spin results in
\begin{equation}\label{cos}
|\vec{p}_{\nu}|\cos{\theta_{\nu}}+|\vec{p}_{e}|\cos{\theta_{e}}+|\vec{p}_{p}|\cos{\theta_{p}}=0,
\end{equation}
 where $\vec{p}_{p}$ is the proton momentum, and ${\theta_{\nu}}$, ${\theta_{e}}$,
 ${\theta_{p}}$ are angles between neutron spin and directions of anti-neutrino,
 electron and proton momenta, correspondingly. From Eq. (\ref{cos}) one can
 see that protons are going to the upper hemisphere
 ($\cos{\theta_{p}}>0$), if
 $|\vec{p}_{\nu}|\cos{\theta_{\nu}}+|\vec{p}_{e}|\cos{\theta_{e}}<0$,
 and to the lower hemisphere
 ($\cos{\theta_{p}}<0$), if
 $|\vec{p}_{\nu}|\cos{\theta_{\nu}}+|\vec{p}_{e}|\cos{\theta_{e}}>0$.
 Therefore,  the $C$-coefficient, being a normalized  difference of the neutron
 decay  rate integrated over neutrino and electron angles, must be
 integrated  over
 the electron energy under these two conditions.
  The integration over azimuthal angles leads to the $4\pi^2$
 factors. To calculate integrals over $\theta_{\nu}$ and $\theta_{e}$,  it is convenient
 to work in cos-variables: $\cos \theta_{\nu}$ and $\cos \theta_{e}$. Thus,
 these two integrals could be represented in terms of a two-dimensional integral in ($\cos \theta_{\nu}$,
 $\cos \theta_{e}$) space, which must be taken separately over lower and upper parts of the
 square area in the cosine plane:  ([-1,1],[-1,1]). The line,  dividing the  area in two
 parts, is given by the equation:  $|\vec{p}_{\nu}|\cos{\theta_{\nu}}+|\vec{p}_{e}|\cos{\theta_{e}}=0$.
 It should be noted that for both these integrals there are  two different
 regimes of integration: $|\vec{p}_{\nu}|>|\vec{p}_{e}|$ and
 $|\vec{p}_{\nu}|<|\vec{p}_{e}|$. For the first case, the integrals
 should be taken first over $\cos{\theta_{\nu}}$ and then over
 $\cos{\theta_{e}}$, and for the second one in the opposite order.
 Applying this  procedure for the decay rate given by Eq.(\ref{cor}), one obtains
 \begin{equation}\label{Ctree}
    C=\frac{X_1}{2X}(A+B)=\frac{X_1}{2X}\frac{4\lambda}{1+3\lambda^2},
  \end{equation}
where
\begin{eqnarray}
 \nonumber
  X &=&
  4\sqrt{(E^{max}_e)^2-m_e^2}(2(E^{max}_e)^4-9(E^{max}_e)^2m_e^2-8m_e^4)\\
  \nonumber
  &+&60E^{max}_em_e^4\ln{((E^{max}_e+\sqrt{(E^{max}_e)^2-m_e^2})
 /m_e)} \\
   \nonumber
  X_1 &=&
  5((E^{max}_e)^5)-6(E^{max}_e)^3m_e^2+3E^{max}_e m_e^4+2m_e^6/E^{max}_e+12E^{max}_em_e^4\ln(E^{max}_e/m_e)).
\end{eqnarray}

 Eq.(\ref{Ctree})   exactly reproduces the results of calculations of
papers \cite{tr58,gl96} (the coefficient in Eq.(\ref{Ctree}) has a
different sign since we define the positive direction of recoil
protons as the direction of the neutron spin polarization). To
obtain a general expression with radiative and recoil corrections,
one has to apply the same procedure for the  general neutron decay
rate given by  Eqs. (8)-(19) in  paper \cite{eftcor}. These
calculations are rather cumbersome but can be done exactly, without
any approximation. Then, one can represent all corrections to the
$C$-coefficient in  Eq.(\ref{Ctree}) as a sum of three terms
\begin{equation}\label{dcparts}
  \Delta C = \Delta C_{\alpha}+\Delta C_{\delta}+\Delta C_{rec},
\end{equation}
where $\Delta C_{\alpha}$ contains Coulomb and radiative
corrections, which do not depend on the nucleon structure (they are
also known  as the "outer" corrections ), $\Delta C_{\delta}$ is the
part of radiative corrections that is dependent on the nucleon
structure (or the "inner" corrections), and $\Delta C_{rec}$
represents recoil corrections. For  recoil corrections we have
\begin{eqnarray}
\label{crec}
 \nonumber
  \Delta C_{rec} &=& \frac{5}{12m_n X (1+3\lambda^2)} \\
  \nonumber
  &\times &\{ 9\lambda \mu_V [(E^{max}_e)^2-m_e^2) (E^{max}_e)^4-4(E^{max}_e)^2m_e^2+
  3m_e^4+4m_e^4\ln(E^{max}_e/m_e)]\\
  \nonumber
   &+& \lambda [31(E^{max}_e)^6)-117(E^{max}_e)^4m_e^2+279 (E^{max}_e)^2
   m_e^4-211m_e^6+18m_e^8/(E^{max}_e)^2 \\
   \nonumber
   &-&12m_e^4(9(E^{max}_e)^2+11m_e^2)\ln(E^{max}_e/m_e)] \\
   \nonumber
   &+& 3\mu_V [-(E^{max}_e)^4m_e^2-9 (E^{max}_e)^2
   m_e^4+9m_e^6+m_e^8/(E^{max}_e)^2 \\
   \nonumber
   &-&12m_e^4((E^{max}_e)^2+m_e^2)\ln(E^{max}_e/m_e)] \\
   \nonumber
   &+& 3\lambda^2 [-2(E^{max}_e)^6)+17(E^{max}_e)^4m_e^2+9 (E^{max}_e)^2
   m_e^4-25m_e^6+m_e^8/(E^{max}_e)^2 \\
      &-&12m_e^4(5(E^{max}_e)^2+m_e^2)\ln(E^{max}_e/m_e)] \\
      \nonumber
   &-& \frac{6X_1 \lambda }{5 X (1+3\lambda^2)}E^{max}_e\sqrt{(E^{max}_e)^2-m_e^2}[\lambda^2 (52(E^{max}_e)^4-124 (E^{max}_e)^2
   m_e^2+507m_e^4) \\
   \nonumber
   &+&(12(E^{max}_e)^4-4 (E^{max}_e)^2
   m_e^2+277m_e^4)]
      \},
 \end{eqnarray}

and for strong interaction dependent part of the radiative
corrections
\begin{equation}\label{cstr}
    \Delta C_{\delta}=-\frac{(4 \lambda X_1 -X)}{2X(1+3\lambda^2)}\frac{\alpha }{2\pi}e^R_V ,
\end{equation}
where $e^R_V$ is low energy constant (LEC) of the EFT \cite{eftcor}.
The expression for $\Delta C_{\alpha}$ is very long and complicated
to be presented here. However, one observes that all coefficients in
the expressions for $\Delta C$ (including $\Delta C_{\alpha}$)
depend only on the mass of electron and the maximal electron energy.
Therefore, one can re-write these expressions in a simple form (and
without a lost of accuracy)  by replacing the mass of electron and
the maximal electron energy with their values: $m_e=0.511099 \; MeV$
and $E_e^{max}=1.293332\; MeV$. Then all dependencies on these
parameters collapse to numerical coefficients in the front of
neutron decay variables and the complete set of corrections $\Delta
C$ could be written as:
\begin{eqnarray}
\label{Ccorrec}
   \Delta C &=&  \frac{1}{(1+3\lambda^2)}[ \frac{\alpha}{2\pi}(23.19375\lambda+4.45619\lambda^2)+\frac{\alpha}{2\pi}e^R_V (0.2748- 1.0993\lambda ) \\
  \nonumber
   &+&
   \frac{1}{m_n}(2.25672\lambda-0.265737\lambda^2-0.0113986\mu_V-0.583714\lambda\mu_V)\\
   \nonumber
   &-& \frac{1}{m_n} \frac{\lambda}{(1+3\lambda^2)}(3.1326+7.775\lambda^2)],
   \label{numcor}
\end{eqnarray}
where neutron mass $m_n$ is in $MeV$.  The first term in the first
line of the Eq.(\ref{numcor}) is $\Delta C_{\alpha}$, the second
term is  $\Delta C_{\delta}$, and last two lines are recoil
corrections. Now, using $m_n=939.57 \; MeV$, $\mu_V=3.7$,
$\alpha=1/137.036$, and $\lambda=1.2695$, one obtains
\begin{equation}\label{totcor}
 \Delta C = 0.0065-0.00022 e^R_V.
\end{equation}
Thus, all radiative and recoil corrections are expressed in terms of
only one unknown parameter (the EFT low energy constant) - which is
supposed to be obtained from another independent experiment, if
possible, or should be calculated from basic principles (for
example, in lattice QCD). In the framework of the EFT, it could be
estimated as $e^R_V \simeq 20$ (see for details \cite{eftcor}).
Discussions of another way of the estimation of $e^R_V $  and its
accuracy is given in the last section.

\section{Neutron $\beta$-decay beyond the Standard model}

Now, when we understand all contributions to $C$ angular correlation
from the Standard model, we can consider how possible contributions
from new physics can change the value of the $C$ asymmetry. To
calculate the possible contributions to the $C$-coefficient from the
models beyond the Standard model, one can use the most general form
of the Hamiltonian for the description of neutron $\beta$-decay in
terms of low energy coupling constants
 $C_i$ (do not confuse with $C$ angular correlation coefficient) by \cite{ly56,gtw1}
 \begin{eqnarray}
 H_{int}&=&(\hat{\psi}_p\psi_n)(C_S\hat{\psi}_e\psi_{\nu}+C^\prime_S\hat{\psi}_e\gamma_5\psi_{\nu})\nonumber \\
&+&(\hat{\psi}_p\gamma_{\mu}\psi_n)(C_V\hat{\psi}_e\gamma_{\mu}\psi_{\nu}+C^\prime_V\hat{\psi}_e\gamma_{\mu}\gamma_5\psi_{\nu})\nonumber \\
&+&\frac{1}{2}(\hat{\psi}_p\sigma_{\lambda\mu}\psi_n)(C_T\hat{\psi}_e\sigma_{\lambda\mu}\psi_{\nu}+C^\prime_T\hat{\psi}_e\sigma_{\lambda\mu}\gamma_5\psi_{\nu})\nonumber \\
&-&(\hat{\psi}_p\gamma_{\mu}\gamma_5\psi_n)(C_A\hat{\psi}_e\gamma_{\mu}\gamma_5\psi_{\nu}+C^\prime_A\hat{\psi}_e\gamma_{\mu}\psi_{\nu})\nonumber \\
&+&(\hat{\psi}_p\gamma_5\psi_n)(C_P\hat{\psi}_e\gamma_5\psi_{\nu}+C^\prime_P\hat{\psi}_e\psi_{\nu})  \label{ham} \\
&+& \text{Hermitian conjugate}, \nonumber
\end{eqnarray}
where the index $i=V$, $A$, $S$, $T$ and $P$ corresponds to vector,
axial-vector, scalar, tensor and pseudoscalar nucleon interactions.
In this presentation, the constants $C_i$ can be considered as
effective constants of nucleon interactions with defined Lorentz
structure, assuming that all high energy degrees of freedom (for the
Standard model and any given extension of the Standard model) are
integrated out. Since we are interested in $C$ angular correlation
coefficient, which is the
 time reversal conserving one, all
constants $C_i$ can be chosen to be real.

 To  see explicitly the influence of a non-standard interactions  on the $C$ angular
 coefficient,
we will follow the procedure described in  paper \cite{ggc}. First,
we re-write the coupling constants $C_i$ as a sum of a contribution
from the standard model $C^{SM}_i$ and a possible contribution from
new physics $\delta C_i$:
 \begin{eqnarray}
C_V &=& C^{SM}_V + \delta C_V \nonumber \\
C^\prime_V &=& C^{SM}_V + \delta C^\prime_V \nonumber \\
C_A &=& C^{SM}_A + \delta C_A \nonumber \\
C^\prime_A &=& C^{SM}_A + \delta C^\prime_A \nonumber \\
C_S &=&  \delta C_S \nonumber \\
C^\prime_S &=&  \delta C^\prime_S \nonumber \\
C_T &=&  \delta C_T \nonumber \\
C^\prime_T &=&  \delta C^\prime_T. \label{consts}
\end{eqnarray}
The pseudoscalar coupling constants are neglected here, since we
treat \cite{gtw1} nucleons nonrelativistically. Then, we  apply the
described above procedure to the calculation of the $C$ angular
correlation coefficient from the Hamiltonian (\ref{ham})  using
Eq.(5) of  paper \cite{ggc}. (It should be noted that in the case of
all $\delta C_i$ being equal to zero, the results is  Eq.
(\ref{Ctree}).) The obtained corrections to the $C$ correlation
coefficient due to contributions from non-standard modes can be
written as:
\begin{eqnarray}
 \nonumber
  \delta C_{NewPhys} &=& \frac{X_1L_1}{2X(1+3\lambda^2)}+\frac{X_3L_3}{2X(1+3\lambda^2)} \\
   &-&
   \frac{X_12\lambda}{X(1+3\lambda^2)}\left[\frac{L_0}{(1+3\lambda^2)}+\frac{X_2L_2}{X(1+3\lambda^2)}\right],
  \label{newphc}
\end{eqnarray}
where
\begin{eqnarray}
 \nonumber
  X_2 &=&
  10m_e(E^{max}_e)\sqrt{(E^{max}_e)^2-m_e^2}(2(E^{max}_e)^2+13m_e^2)\\
  \nonumber
  &-&30m_e (4 (E^{max}_e)^2m_e^2+m_e^4)\ln{((E^{max}_e+sqrt{(E^{max}_e)^2-m_e^2})
 /m_e)} \\
   \nonumber
  X_3 &=&
  5m_e(3(E^{max}_e)^4+12(E^{max}_e)^2m_e^2-15 m_e^4-12(2(E^{max}_e)^2m_e^2 +m_e^4)\ln(E^{max}_e/m_e)).
\end{eqnarray}
The coefficients $L_i$  depend only on new physics contributions:
\begin{eqnarray}
 \nonumber
  L_0 &=&  (\delta C_V+\delta C^\prime_V )+ ({\delta C_V}^2+{\delta C^\prime_V}^2+{\delta C_S}^2+{\delta C^\prime_S}^2)/2 \nonumber \\
 &+& 3 [ \lambda (\delta C_A +\delta C^\prime_A)+ ({\delta C_A}^2+{\delta C^\prime_A}^2+{\delta C_T}^2+{\delta C^\prime_T}^2)/2],  \\
  \nonumber
  L_1 &=& -2(\delta C_A+{\delta C^\prime_A})  +3 \delta C_T \delta C^\prime_T -2 (\delta C_V \delta C^\prime_A +\delta C^\prime_V \delta C_A) +2 \lambda(\delta C_V+\delta C^\prime_V ), \\
  L_2 &=& \sqrt{1-\alpha^2}[(\delta C_S+\delta C^\prime_S )+\delta C_S \delta C_V+ \delta C^\prime_S \delta C^\prime_V \nonumber \\
&+& 3(\lambda(\delta C_T +\delta C^\prime_T)+\delta C_T \delta C_A+ \delta C^\prime_T \delta C^\prime_A )], \\
  \nonumber
    L_3 &=&  \sqrt{1-\alpha^2}[-2\lambda (\delta C_T+\delta C^\prime_T)-\lambda (\delta C_S+\delta C^{\prime}_S) +(\delta C_T+C^\prime_T)  \nonumber \\
 &+& 2 \delta C_T \delta C^\prime_A +2 \delta C_A \delta C^\prime_T +\delta C_S \delta C^\prime_A +\delta C_A \delta C^\prime_S + \delta C_V \delta C^\prime_T +\delta C_T \delta C^\prime_V]
  \label{Lcoef}
\end{eqnarray}
In the above expressions,  we have neglected radiative corrections
and recoil effects for the new physics contributions, but kept
Coulomb corrections since they can be important for a low energy
part of the electron spectrum.

From Eq.(\ref{newphc}), one can see that, as in the case of
radiative and recoil corrections, all coefficients in the expression
 are functions only of electron mass and maximum
electron energy. Therefore, we simplify the general expressions for
the contributions from new physics, by substituting numerical values
for all known parameters (electron mass, electron maximal energy, as
well as for $\alpha=1/137.036$ and $\lambda=1.2695$) and  keep only
first order contributions from non-standard interactions. Then,
Eq.(\ref{newphc}) transforms into
\begin{eqnarray}\label{NumNPC}
     \delta C_{NewPhys}&=&0.05657 (\delta C_V+\delta C^\prime_V
     )+0.04456 (\delta C_A+\delta C^\prime_A )\nonumber  \\
     &-&0.06234 (\delta C_S+\delta C^\prime_S
     )+0.02132 (\delta C_T+\delta C^\prime_T ).
\end{eqnarray}

Instead of the presentation of these corrections in terms of low
energy coupling constants related
 to the Lorentz structure of weak interactions, we can re-write them
 in terms of quark and lepton current constants $\bar{a}_{jl}$ and $\bar{A}_{jl}$, defined in  paper \cite{herc}.
   Using the transformation rules \cite{ggc} :
\begin{eqnarray}
\delta C_V +\delta C^\prime_V&=& 2 (\bar{a}_{LL}+\bar{a}_{LR}), \nonumber  \\
\delta C_A +\delta C^\prime_A &=& 2\lambda (\bar{a}_{LL}-\bar{a}_{LR}),  \nonumber \\
\delta C_S +\delta C^\prime_S&=& 2g_S (\bar{A}_{LL}+\bar{A}_{LR}),  \nonumber \\
\delta C_T+\delta C^\prime_T &=& 4 g_T \bar{\alpha}_{LL},
 \label{carel}
\end{eqnarray}

and assuming \cite{herc} $g_S=1$ and $g_T=1$,  we obtain the
expression for corrections from new physics as:
\begin{eqnarray}\label{NumNPa}
     \delta C_{NewPhys}&=&0.11314 (\bar{a}_{LL}+\bar{a}_{LR}
     )+0.11314 (\bar{a}_{LL}-\bar{a}_{LR} )\nonumber  \\
     &-&0.12468 (\bar{A}_{LL}+\bar{A}_{LR}
     )+0.08528 \bar{\alpha}_{LL}.
\end{eqnarray}

The parameters $\bar{a}_{jl}$, $\bar{\alpha}_{jl}$, and
$\bar{A}_{jl}$  describe contributions to the low energy Hamiltonian
from current-current interactions in terms of $j$-type of leptonic
current and $i$-type of quark current. For example, $\bar{a}_{LR}$
is the contribution to the Hamiltonian from left-handed leptonic
current and right-handed quark current normalized by the size of the
Standard Model (left--left current) interactions.
 $g_S$ and $g_T$ are formfactors at zero-momentum transfer in the nucleon matrix element of scalar and tensor currents.
  For more details, see  paper \cite{herc}.

 The expected values of these parameters vary over a wide range from $0.07$ to $10^{-6}$ (see Table \ref{nptab}
 and paper \cite{herc}
 for the comprehensive analysis and for  discussions of significance of each of these
parameters for models beyond the Standard one).

\begin{table}
  \centering
  \caption{Possible manifestations of new physics}
  \label{nptab}
\begin{tabular}{|c|c|c|c|c|c|c|}
  \hline
Model & L-R & Exotic Fermion & Leptoquark & Contact interactions & SUSY & Higgs \\
\hline
$\bar{a}_{LL}$ &   &   & 0.2 - 0.03 &  &  &  \\
$\bar{a}_{LR}$ &   & 0.01 & 0.01 &   &   &   \\
$\bar{A}_{LL}+\bar{A}_{LR}$ &   &   &  & 0.01 & $7.5 \cdot 10^{-4}$ & $3\cdot 10^{-6}$ \\
$-\bar{A}_{LL}+\bar{A}_{LR}$ &   &   & $3\cdot 10^{-6}$ &   &   &  \\
\hline\end{tabular}
\end{table}

\section{Conclusions}

Taking into account the results of Eqs.(\ref{Ctree}),
(\ref{dcparts}), (\ref{newphc}), and (\ref{NumNPa}), one can write
the complete expression for the $C$ angular coefficient
($C_{total}$) as a sum of the tree-level approximation $C$,
radiative and recoil corrections in the Standard Model $\Delta C$,
and possible contributions from new physics $\delta C_{NewPhys}$:
\begin{equation}\label{cgen}
    C_{total}=C+\delta C +\delta C_{NewPhys}.
\end{equation}
It should be noted, that this equation is the exact expression of
the $C$ angular correlation coefficient in the first order of recoil
corrections, radiative corrections, and low energy contributions
from new physics. Therefore, it could be considered as the complete
expression up to the level of  accuracy of $10^{-5}$, provided the
EFT low energy constant (LEC) is given. Otherwise, it could be
considered as a parametrization in terms of one free parameter -
 LEC with the same accuracy of $10^{-5}$. Would the parameter $e^R_V$ be determined
from another independent experiment (for example, from the precise
measurement of neutrino-deuteron cross-sections) or calculated using
lattice QCD approach,  Eq. (\ref{cgen}) could be used to test the
Standard model  up to the level of accuracy  of about $10^{-5}$, by
comparing a theoretical prediction with experimental results.
Unfortunately,  neutrino experiments and QCD calculations with the
required accuracy are rather difficult problems and we cannot rely
on them at the present time.

To understand the desirable level of  accuracy  in a search for new
physics, one can use first a conservative approach: the estimate for
the LEC as $e^R_V \simeq 20$ given in  paper \cite{eftcor}. Then,
the level of theoretical uncertainties due to strong interactions,
according to Eq.(\ref{totcor}), is about $0.0044$, which is
comparable to the claimed experimental accuracy $0.0026$ of the
recent experiment \cite{CILL}. However, as it was mentioned in
\cite{eftcor} that by comparing the results of the EFT approach and
the calculations of radiative corrections for total neutron decay
rate \cite{sirlin,sirlinrmp,sir}, one can find the correspondence
between these two calculations, which results \cite{eftcor} in the
following equation
\begin{equation}\label{LEC}
e_V^R = -\frac{5}{4} -4\, {\rm ln}\left(\frac{m_W}{m_Z}\right)
+3\, {\rm ln}\left(\frac{m_W}{m_N}\right) + {\rm
ln}\left(\frac{m_W}{m_A}\right) + 2 C_{Born} + A_g  .
\end{equation}
Here $m_W, m_Z$ are the masses of the W, Z bosons and $m_A$ is the
axial mass scale, which are rather well known. The source of
theoretical uncertainties is related to two last terms $C_{Born}$
and $A_g$ (see, for details \cite{sirlin,sirlinrmp,sir,marc06}).
Changing from the EFT "ideology" with one unknown LEC  to  direct
calculations using strong interaction models,  we lost the
attractive feature of the model independent EFT approach  and have
to deal with dependencies on strong interaction models applied for
description internal structure of nucleons. On the other hand, in
the given framework \cite{sirlin,sirlinrmp,sir}, which is actually a
very well recognized  standard approach to general analysis of weak
interactions, we can reduce
 uncertainties in the estimation of LEC to the uncertainties of calculations
of $C_{Born}$ and $A_g$ terms. Then, using results of recent
calculations of these terms \cite{marc06} $C_{Born}\simeq 0.829$ and
$A_g\simeq -0.34$ with the claimed level of uncertainty of $10\%$,
one can reduce the level of uncertainty of the obtained theoretical
description of the $C$ angular coefficient to the level of about
$10^{-5}$, i. e. to the level  of validity of the description of
neutron decay in  paper \cite{eftcor}.

Accepting these estimates, one can see from Eqs. (\ref{NumNPC}) and
(\ref{NumNPa}) that precise measurements of the $C$ angular
correlation can provide limits for non-standard interactions in
terms of $\delta C_i$ coupling constants up to the level of about
$(2-5)\cdot 10^{-4}$, or, in terms of parameters related to
non-standard currents, up to the level of about $ 10^{-4}$. However,
in order to be able to constrain new physics parameters at this
level,  the currently achieved experimental accuracy \cite{CILL}
must be improved by two orders of magnitude.


\begin{acknowledgments}
I thank T. Chupp, who brought this problem to my attention .
 This work was supported by the DOE grant no. DE-FG02-03ER46043.
\end{acknowledgments}

\end{document}